\documentclass{aa}
\usepackage{graphicx}

\begin{document}

\thesaurus{06(08.03.4;08.05.2;08.09.2 4U\,0115+63;08.02.1;08.14.1;13.25.5)} 

\title{The Be/X-ray transient 4U\,0115+63/V635 Cas}
\subtitle{I. A consistent model}

\author{Ignacio~Negueruela\inst{1,2}\thanks{Now at Observatoire
de Strasbourg, 11 rue de l'Universit\'{e}, F67000 Strasbourg, France}
\and Atsuo~T.~Okazaki\inst{3}}                   
                                                            
\institute{SAX SDC, ASI, c/o Nuova Telespazio, via Corcolle 19, I00131
Rome, Italy
\and Astrophysics Research Institute, Liverpool John Moores University, 
Byrom St., Liverpool, L3 3AF, UK
\and Faculty of Engineering, Hokkai-Gakuen University,Toyohira-ku, Sapporo
062-8605, Japan}

\mail{ignacio@astro.u-strasbg.fr}

\date{Received    / Accepted     }

\titlerunning{A model for 4U\,0115+63}
\authorrunning{Negueruela \& Okazaki}
\maketitle 

\begin{abstract}

We present photometry and high SNR spectroscopy in the classification 
region of V635~Cas, the optical counterpart to the transient X-ray 
pulsator 4U\,0115+63, taken at a time when the circumstellar envelope
had disappeared. V635~Cas is classified as a B0.2Ve star at a 
distance of $7-8$ kpc. We use the physical parameters derived from
these observations and the orbit derived from X-ray observations to 
elaborate a model of the system based on the theory of decretion discs
around Be stars. We show that the disc surrounding the Be star must be
truncated by the tidal/resonant interaction with the neutron star and
cannot be in a steady state. This explains many of the observed
properties of 4U\,0115+63. In particular, because of this effect, 
under normal circumstances, the neutron star cannot accrete from the
disc, which explains the lack of regular Type~I outbursts from the source.
 \end{abstract}

\keywords{stars: circumstellar matter -- emission-line, Be -- individual: 
4U\,0115+63, -- binaries:close -- neutron   -- X-ray: stars

\section{Introduction}

The hard X-ray transient 4U\,0115+63 (X\,0115+634) is one of the best 
studied  Be/X-ray
binary systems (see Campana 1996; Negueruela et al. 1997, henceforth 
N97). More than 50 of these systems, in which a neutron star orbits a Be 
star in a moderately eccentric orbit, are known (see Negueruela 1998; 
Bildsten et al. 1997). The 
Be star is surrounded by a disc of relatively cool
material, presumably ejected from the star due to causes unknown, but
generally believed to be associated with fast rotation, magnetic fields
and/or non-radial
pulsations (Slettebak 1988). The presence of the disc gives rise to  
emission lines in the optical and infrared spectral regions and an 
excess in the infrared continuum radiation.

The hard X-ray emission is due to the accretion of circumstellar material 
on to the neutron star companion. Due to their different geometries and the 
varying physical conditions in the circumstellar disc, Be/X-ray binaries 
can present very different states of X-ray activity (Stella et al. 1986). 
In quiescence, they display persistent low-luminosity 
($L_{{\rm x}} \la 10^{36}$ erg s$^{-1}$) X-ray emission or no detectable 
emission at all. Occasionally, they show 
series of periodical (Type I) X-ray outbursts ($L_{{\rm x}} 
\approx 10^{36} - 10^{37}$ erg s$^{-1}$), separated by the orbital period
of the neutron star. More rarely, 
they undergo giant (Type II) X-ray outbursts ($L_{{\rm x}} \ga 10^{37}$ 
erg s$^{-1}$), which do not clearly correlate with the orbital motion. 
Some systems only display persistent emission, but most of them show 
outbursts and are termed Be/X-ray transients.

\begin{figure*}
\begin{picture}(500,250)
\put(0,0){\includegraphics{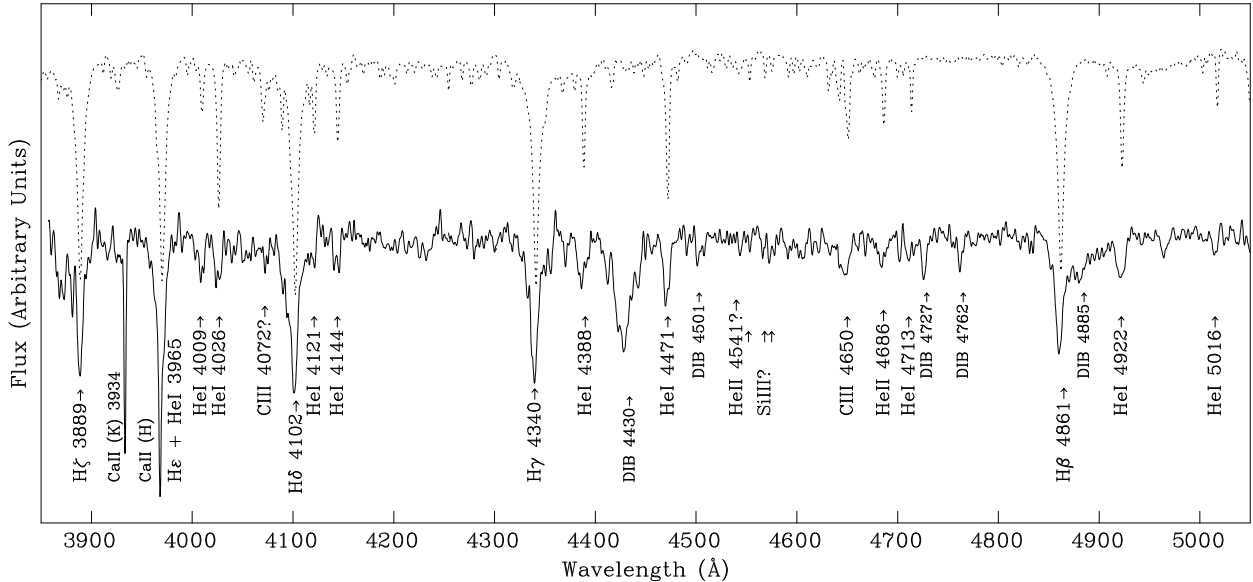}}
\end{picture}
\caption{The spectrum of V635~Cas in the classification region. Two 
exposures taken on November 14, 
1997, with ISIS on the WHT equipped with the R1200B 
grating and the EEV10 camera have been combined for this figure.
The comparison spectrum is that of the B0.2V standard $\tau$ Sco. 
Both spectra have been divided by a spline fit to the continuum for
normalisation and smoothed with a $\sigma
= 0.8$\AA\ Gaussian function for display.} 
\label{fig:bluespec} 
\end{figure*}

The transient 4U\,0115+63 was first reported in the {\em Uhuru}
satellite survey (Giacconi et al. 1972; Forman et al. 1978), though a search 
of the {\em Vela 5B} data base revealed that the source had already been 
observed by this satellite since 1969 (Whitlock et al. 1989). Precise 
positional determinations by the {\em SAS 3}, {\em Ariel V} and 
{\em HEAO-1} satellites (Cominsky et al. 1978; Johnston et al. 1978) were 
used to identify the system 
with a heavily reddened Be star with a visual magnitude $V \approx 15.5$ 
(Johns et al. 1978; Hutchings \& Crampton 1981), which was subsequently 
named V635~Cas (Khopolov et al. 1981). Rappaport et al. (1978) used 
{\em SAS 3} timing observations to derive the orbital parameters of the 
binary system, which consists of a fast-rotating ($P_{{\rm s}} = 3.6 \: 
{\rm s}$) neutron star in a relatively close ($P_{{\rm orb}} = 
24.3 \: {\rm d}$) and eccentric ($e = 0.34$) 
orbit around the Be star (see also Tamura et al. 1992). 

Due to the fast rotation of the neutron star, centrifugal inhibition of 
accretion prevents the onset of X-ray emission unless the ram pressure
of accreted material reaches a relatively high value (Stella et al. 1986;
N97). The system 
had only be known to display Type II activity until a short series of Type I 
outbursts was detected by BATSE and {\em RXTE} in 1996 (Bildsten et al. 1997; 
Negueruela et al. 1998). The giant outbursts are associated with large
amplitude brightenings of the optical and infrared magnitudes of the 
counterpart (N97).

This is the first of two papers dedicated to providing a coherent picture
of this system and understanding the implications of its unusual behaviour
for the general class of Be/X-ray transients. Here we derive the 
astrophysical parameters of 4U\,0115+63 and build a model for circumstellar 
disc around the Be star that allows us to understand the usual
quiescent state of the system. In the second paper
(Negueruela et al. 2000; henceforth Paper II), we will analyse the temporal
evolution of this disc and investigate how its behaviour is connected
with the X-ray activity of the source.

\section{Observations}

We present optical photometry and optical
spectroscopy of V635~Cas, obtained, for the first time, 
when the emission component was almost completely absent.

\subsection{Optical spectroscopy}

Observations of the source were taken  on November 14, 1997, using the 
Intermediate Dispersion Spectroscopic and Imaging System (ISIS) on the 
4.2-m William Herschel Telescope (WHT), located at the Observatorio del 
Roque de los Muchachos, La Palma, Spain. The 
blue arm was equipped with the R1200B grating and the EEV10 CCD, which
gives a nominal dispersion of $\sim 0.22$ \AA/pixel over $\sim 900$ \AA.
The resolution at $\sim \lambda$4600\AA, estimated from the FWHM of arc 
lines, is $\sim 0.7$ \AA.
Two exposures were taken, centred on $\lambda$4250\AA\ and $\lambda$4650\AA.
A combined spectrum is shown in Fig.~\ref{fig:bluespec}.

The red arm was equipped with the R1200R grating and the Tek5 CCD,
which gives a nominal dispersion of $\sim 0.4$ \AA/pixel at H$\alpha$
(the resolution is $\sim 0.8$\AA\ at H$\alpha$).
A similar observation has been attempted on July 19, 1997, but due to bad
weather conditions, the blue spectrum is too noisy to be of any use. 
An H$\alpha$ spectrum, though, was obtained using the same setting as
in the November observations. The observations are displayed in 
Fig.~\ref{fig:red}.
All the data have been reduced using the {\em Starlink}
software packages {\sc ccdpack} (Draper 1998) and {\sc figaro} 
(Shortridge et al. 1997) and
analysed using {\sc figaro} and {\sc dipso} (Howarth et al. 1997).

\subsection{Optical photometry}

\begin{figure}[ht]
\begin{picture}(250,150)
\put(0,0){\includegraphics{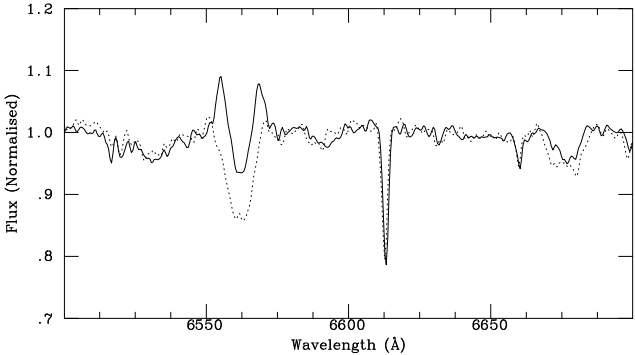}}
\end{picture}
\caption{High resolution spectra of V635~Cas showing the almost completely 
disc-less state observed in 1997. In the spectrum from July 19 (dashed line)
only some residual emission is visible in the \ion{He}{i} $\lambda$6678 \AA\ 
line. In the November 14 spectrum, taken simultaneously with the blue 
spectroscopy, the disc is beginning to reform and some emission is present 
now in H$\alpha$. The peak separation corresponds to 
$\Delta v_{{\rm peak}} \approx 600 \: {\rm km}\,{\rm s}^{-1}$,
well above what would be seen in the case of angular momentum conservation.
The strong narrow features are interstellar diffuse bands. The spectra have 
been divided by a spline fit to the continuum for normalisation.}
\label{fig:red}
\end{figure}

From the observations listed in Paper II, we have taken the dataset 
showing the faintest and bluest magnitudes, which corresponds to 1998
January 7. The values measured are $U=17.05\pm0.10$, $B=16.92\pm0.07$, 
$V=15.53\pm0.04$,  $R=14.55\pm0.03$ and $I= 13.49\pm0.03$. As discussed 
in Paper II,
these observations coincided with a disc-less phase of V635~Cas and should
represent values close to the intrinsic magnitudes of the star. We note
that the value for $I$ is compatible with the faintest point measured
by Mendelson \& Mazeh (1991) in the five years that their photometry covers.

\section{Results}

\subsection{Spectral classification}

Unger et al. (1998) have presented intermediate resolution spectroscopy
of V635~Cas and classified the source as O9.5V. However, their spectrum 
had a very low signal to noise ratio and showed several spurious 
features.
The spectrum shown in Fig. \ref{fig:bluespec} has a much higher SNR and 
better resolution. At the time it was taken, H$\alpha$ was starting to 
develop emission peaks after several months during which it had been observed
in absorption (see Paper II). All the other Balmer lines are still 
in absorption, but they all present a similar asymmetric shape, which must 
be due to new emission components starting to develop. Surprisingly, many
He\,{\sc i} lines show emission components ($\lambda\lambda$ 4009,
4026, 4144 and 4713 \AA), as can be easily seen by comparing the spectrum
with that of the B0.2V standard $\tau$ Sco (Fig. \ref{fig:bluespec}).

The broad shallow lines are typical of a main-sequence early star.
He\,{\sc ii} $\lambda$4686 \AA\ can be clearly seen, though it is weak,
indicating a spectrum earlier than B0.7. The weakness of the Si\,{\sc iii}
$\lambda\lambda$ 4552, 4568, 4575 \AA\ triplet (if it is at all present) 
supports the early spectral type and indicates
a main sequence star. No obvious O\,{\sc ii} lines
are present, though the C\,{\sc iii} $\lambda\lambda$ 4072, 4650 \AA\ lines
could be blends.  If the weak feature at $\lambda$4541\AA\ is real, then
V635~Cas could be as early as B0V, but not earlier -- the \ion{Si}{iv}
lines, which must be hidden on the very broad wings of H$\delta$, should 
be clearly visible for an earlier (or higher luminosity) star. Therefore
the star is constrained to belong to the B0-B0.5 range. For simplicity, we 
will adopt the intermediate B0.2Ve spectral type.

\subsection{Distance}
\label{sec:dista}
 
As discussed in Paper II, during the second half of 1997 
the source showed H$\alpha$ and all other lines in the red in absorption. 
On January 7th, 1998, we measured the faintest and bluest magnitudes in 
our dataset. Here we investigate the possibility that these photometric
magnitudes are close to the actual apparent magnitudes of the star without 
any contribution from a disc, implying that the measured reddening is
(almost) purely interstellar. 

The intrinsic colour of a B0V star is $(B-V)_{0}=-0.26$ 
(Wegner 1994) and therefore the measured $(B-V)= 1.39$ 
implies a reddening $E(B-V)=1.65\pm0.08$. Using the relationship 
$E(U-B) = E(B-V) [0.69+0.04\times E(B-V)]$ from 
Fitzpatrick (1999), we find $E(U-B) = 1.25$. The measured 
$(U-B) = 0.13$ implies $(U-B)_{0}=-1.12\pm0.12$, perfectly compatible
with the expected $(U-B)_{0}=-1.08$. Further assuming the standard
extinction curve for $R=3.1$ after Fitzpatrick (1999), we find values
for $(V-R)_{0}=-0.31$ and $(V-I)_{0}= -0.60$. These values are, within
the errors, close to the intrinsic colours of a B0V star, though 
slightly too blue. Wegner (1994) gives $(V-R)_{0}=-0.14$ and 
$(V-I)_{0}= -0.37$. Such an effect is surprising, since any residual
emission is expected to contribute more strongly at longer wavelengths.

\begin{figure*}[ht]
\begin{picture}(500,280)
\put(0,0){\includegraphics{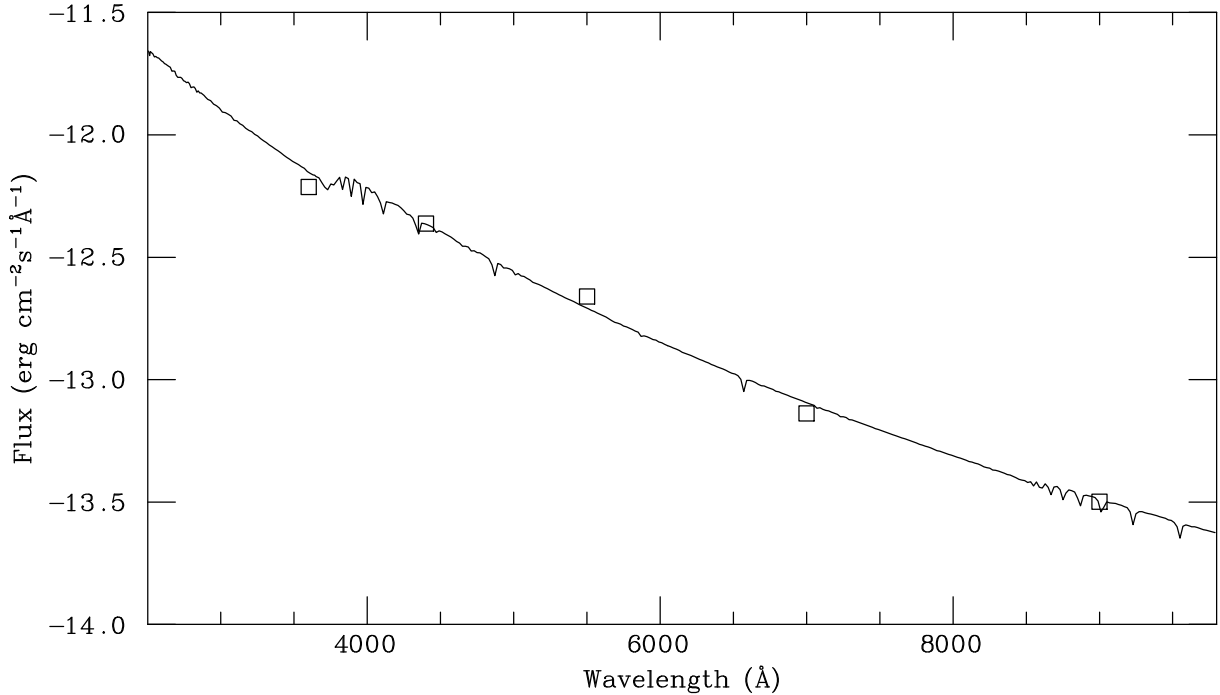}}
\end{picture}
\caption{A Kurucz model for $T_{\rm eff} = 26000\:{\rm K}$ and 
$\log g = 4.0$ (the adopted parameters for V635~Cas) has been 
normalised to the $B$-band magnitude and is compared to the
measured energy distribution (points have been set at the effective
wavelengths of the different bands) dereddened with $R=3.1$ and
$E(B-V)=1.60$ (a value close to the lower limit allowed by the error
bars in our photometry).}
\label{fig:specdist}
\end{figure*}

We have considered the possibility that the extinction law to V635~Cas 
is different from the standard, since the line of sight to V635~Cas 
could intersect the outer regions of the OB association Cas OB7. 
This association is about 3$^{\circ}$ in extension,
centred at $\sim$ ($l=123$, $b=+1$) and the de-reddened $DM=11.3$ implies a
distance of 1.8 kpc (Garmany \& Stencel 1992). We have dereddened the 
photometry using the extinction curves of
Fitzpatrick (1999) for different values of $R$. We find that values of
$R$ different from the standard value by more than $\sim 0.2$ provide 
dereddened energy distributions that are farther away from the theoretical
values than that dereddened with $R=3.1$. On the other hand, the derived 
colours become very close to those expected when
a value of $E(B-V)$ close to the lower limit allowed by the errors is 
considered.

We have also measured the EW of the interstellar diffuse band
at $\lambda$6613\AA\ in 
several H$\alpha$ spectra of V635~Cas (N97; Paper II). We have 
obtained an average value EW = 
480$\pm$100 m\AA. According to the relation by Herbig (1975), the implied 
colour excess is $E(B-V) = 1.9\pm0.4$, consistent with the photometric
value. Likewise, the EW of the interstellar diffuse band at 
$\lambda$4430\AA\ is $4.4\pm0.4$, which results (Herbig 1975) in  
$E(B-V) = 2.0\pm0.2$. We note that the \ion{Na}{i} H and K lines cannot
be used for distance calibration, since the values measured are
well above the saturation value for the relationship with reddening
(Munari \& Zwitter 1997).

Even though the interstellar diffuse bands seem to support a  
value for $E(B-V)$ close to the upper limit allowed by the error bars,
the photometric evidence supports a value close to the lower limit, as can
be seen in Fig.~\ref{fig:specdist}. The choice of a lower $E(B-V)$
results in a higher distance for V635~Cas.

Assuming an absolute magnitude 
$M_{V} = -4.1$ after Vacca et al. (1996), and taking $E(B-V)=1.65\pm0.08$,
we derive a distance to V635~Cas of $8\pm1\:{\rm kpc}$. This could
be an overestimate if some residual emission was still contributing 
to the intrinsic luminosity of V635~Cas. Moreover, the calibration of
Vacca et al. (1996) indicates that early-B stars are hotter than previously
thought and therefore the intrinsic $(B-V)_{0}$ used is probably too low.
We note that distances $\sim 7$ kpc are obtained for the Be/X-ray 
transient V0332+53  (Negueruela et al. 1999) and for the standard 
High Mass X-ray Binary 2S\,0114+65 (Reig et al. 1996), which is very 
close in the sky to 4U\,0115+63. We consider very likely that the
distance to 4U\,0115+63 is also $\approx 7$ kpc and that all these 
objects belong to an outer arm of the Galaxy, situated beyond the 
Perseus arm at a distance $\ga 4$ kpc (Kimeswenger \& Weinberger 1989).
At this distance, the maximum luminosity of the source,
measured by $Vela$ 5$B$ during a Type II outburst in 1974 (Whitlock et al.
1989) is close to $\ga 10^{38}$ erg s$^{-1}$, i.e., close to the
Eddington luminosity for a neutron star.

\section{A model for 4U\,0115+63}

\subsection{Stellar and orbital parameters}
\label{sec:param}

Given the B0.2Ve spectral type of the optical component, we expect it to
have a mass $M_{*} \simeq 19\: M_{\sun}$ (Vacca et al. 1996). In this case, 
the mass function $f(M)=5.0\: M_{\sun}$ implies an inclination angle 
$i=42^{\circ}$ for a standard neutron star mass $M_{{\rm x}}= 1.4\: M_{\sun}$.
It is noteworthy that making the companion star moderately undermassive 
will not change strongly the inclination angle. For example, if $M_{*} 
= 12\: M_{\sun}$, the inclination angle is $i=54^{\circ}$. For any 
reasonable value of $M_{*}$, $40^{\circ} \la i \la 60^{\circ}$ or $0.7 \la 
\sin i \la  0.8$. The dependence on $M_{{\rm x}}$ is still smaller.
Therefore, in what follows we will adopt as a model
$M_{*} = 18\, M_{\sun}$ and $M_{{\rm x}}= 1.4\, M_{\sun}$, implying 
$i=43^{\circ}$. In this case, the measured $a_{{\rm x}}\sin i$ implies
a binary separation $a=6.6\times10^{10} \: {\rm m} = 95\: R_{\sun}$. Assuming 
$R_{*} = 8\: R_{\sun}$ (Vacca et al. 1996), this results in periastron
and apastron distances $a_{{\rm per}} = 8\: R_{*}$ and $a_{{\rm ap}} =
16\: R_{*}$.

We have tried to estimate the inclination of the equatorial plane of V635
Cas by using Buscombe's (1969) approximation and the fits to the FWHM of 
He\,{\sc i} $\lambda \lambda$ 4388, 4471 \AA\ of Steele et al. (1999). 
This is only a coarse approximation, since weak emission components are
present on the wings of the lines and, given the relatively low SNR of 
the blue spectrum and the broadness of the Balmer lines, several weak lines 
could be contaminating the line shapes. In spite of
this, all the measurements obtained fall in the range $v \sin i 
\approx 240 - 340$ km s$^{-1}$ with most of them concentrated around
$v \sin i = 290$ km s$^{-1}$ and the Balmer lines consistently giving
higher values than the He\,{\sc i} lines (except H$\alpha$).

Therefore, allowing for the effect
of emission contamination, we will adopt as an estimate 
$v \sin i \approx 300$ km s$^{-1}$, taking into account that the value is not
likely to be much smaller, but could be larger. We note
that, if we assume coplanarity between the equatorial and orbital plane,
this value implies a rotational velocity $v=440$ km s$^{-1}$, which is 
compatible with the assumption that Be stars rotate at $\la 80\%$ of 
their break-up velocity. Moreover, the double-peaked emission lines 
which are seen soon after the circumstellar disc starts reforming
are typical of a moderate inclination 
($i \approx 40^{\circ}\,-\,60^{\circ}$), confirming that the value of
$\sin i$ cannot be very distant.

Therefore, we find no compelling reason to argue for an inclination 
between the orbital and equatorial plane. As
discussed in Paper II, this does not imply that the 
circumstellar disc must always remain coplanar.

\subsection{Viscous disc around V635~Cas}
\label{sec:disc}

To date, discs around seven Be stars have been spatially resolved with
optical interferometers (Quirrenbach et al.\ 1997). The size of the
H$\alpha$-emitting region $R_{{\rm H}\alpha}$ of these stars ranges
from 3 to 12 stellar radii. The separation of the two peaks of
H$\alpha$, $\Delta v_{\rm peak}$, is significantly wider than $2 v
\sin i \,(R_{{\rm H}\alpha}/R_*)^{-1}$ expected in an angular-momentum
conserving disc, but is consistent with a disc structure in which the
rotation velocity is near Keplerian.

  Although there is no widely-accepted
model to form near-Keplerian discs around Be stars, a viscous
decretion disc model proposed by Lee et al.\ (1991) seems
promising (Porter 1999; see also Okazaki 2000a). In this model, the matter 
supplied
from the equatorial surface of the star drifts outward because of the viscous
effect and forms the disc. Basic equations for viscous decretion discs
are the same as those for viscous accretion discs, except that the
sign of $\dot{M}$ (mass decretion/accretion rate) is opposite.
The viscous decretion disc model thus predicts a geometrically thin and 
near-Keplerian disc around a Be star.
The outflow in the viscous decretion disc is highly subsonic
near the star (Okazaki 1997, 2000b), which is consistent with the
observed upper limit on the radial velocity ($3\,{\rm km\,s}^{-1}$)
in discs around Be stars (Hanuschik 2000).

Since the mass of the neutron star is much smaller than that of the Be star, 
its presence will hardly affect the mass loss process from the Be star and 
will have a minor effect on the formation and structure of the Be disc 
except in the region near the critical Roche potential. We note that
the overall characteristics of the discs in Be/X-ray binaries, as
derived from the observation of emission lines and their evolution
are not fundamentally different from those of isolated Be stars 
(see Negueruela et al. 1998, where a discussion of the similarities
and main differences is presented; and see also Section~\ref{sec:discuss}).

Therefore, as a first approximation, we will assume in this paper
that the viscous decretion disc model,
which has been developed as a disc model for isolated Be stars,
is applicable to Be discs in Be/X-ray binaries.
That is, we will assume that the disc surrounding V635~Cas is
near-Keplerian when unperturbed. For simplicity, we also assume that
the disc is isothermal with a temperature of $0.8 T_{\rm eff}$,
where the effective temperature of the star is $T_{\rm
   eff}=26000\,{\rm K}$.
Table~\ref{tbl:params} summarizes the model parameters adopted for
4U\,0115+63.

\begin{table}[t]
    \caption{Model parameters for V635~Cas/4U\,0115+63.}
    \begin{center}
       \begin{tabular}{llcc}
          \hline
          \multicolumn{2}{c}{Parameter} &
          \multicolumn{1}{c}{Notation} &
          \multicolumn{1}{c}{Value} \\
          \hline
          \multicolumn{4}{l}{Binary orbit} \\
          & Period & $P_{\rm orb}$ & 24.3\,d \\
          & Inclination & $i$ & $43^\circ$ \\
          & Semi-major axis & $a$ & $95\,R_{\sun}(=12\,R_{*})$ \\
          & Eccentricity & $e$ & 0.34 \\
          \multicolumn{4}{l}{Neutron star} \\
          & Mass & $M_{\rm x}$ & $1.4\,M_{\sun}$ \\
          \multicolumn{4}{l}{Be star} \\
          & Mass & $M_*$ & $18.0\,M_{\sun}$ \\
          & Radius & $R_*$ & $8.0\,R_{\sun}$ \\
          & Effective temperature & $T_{\rm eff}$ & 26,000\,K \\
          \multicolumn{4}{l}{Be disc} \\
          & Temperature & $T_{\rm d}$ & 20,800\,K \\
          & Scale-height at $r=R_*$ & $H(R_*)/R_*$ & $2.6 \cdot 10^{-2}$ \\
          \hline
       \end{tabular}
       \label{tbl:params}
    \end{center}
\end{table}

\subsection{Tidal truncation of the viscous disc}
\label{sec:trunc}

Artymowicz \& Lubow (1994) investigated the tidal/resonant truncation
of the accretion discs in eccentric binary systems and found that the
disc size becomes smaller in a system with larger
eccentricity. Following their formulation of the companion-disc
interaction, we study below the tidal truncation of the Be-star disc
in 4U\,0115+63.

In order to evaluate the tidal/resonant torque exerted by the neutron star
on the Be-star disc, we consider the binary potential in a
coordinate system $(r, \theta, z)$ in which the origin is attached
to the Be star primary:
\begin{eqnarray}
    \Phi (r, \theta, z) &=& -{{GM_{*}} \over r}
    -{{GM_{\rm x}} \over{[r^2+r_2^2-2rr_2 \cos(\theta-f)]^{1/2}}} \nonumber \\
    && +{{GM_{\rm x}r} \over r_2^2} \cos(\theta-f),
    \label{eqn:pot1}
\end{eqnarray}
where $M_{*}$ and $M_{\rm x}$ are masses of the Be and neutron stars,
respectively, $r_2$ is the distance of the neutron star from the
primary, and $f$ is the true anomaly of the neutron star. The third
term in the right hand side of Eq.(\ref{eqn:pot1}) is the indirect
potential arising from the fact that the coordinate origin is at the primary.

We expand the potential by a double series as
\begin{equation}
    \Phi (r, \theta, z)
    = \sum_{m,l} \phi_{ml} \exp[i(m \theta - l \Omega_B t)],
    \label{eqn:pot2}
\end{equation}
where $m$ and $l$ are the azimuthal and time-harmonic numbers,
respectively, and $\Omega_B = [G(M_{*}+M_{\rm x})/a^3]^{1/2}$ is the mean
motion of the binary with semimajor axis $a$. The pattern speed of
each potential component is given by $\Omega_p=(l/m)\Omega_B$.

Inverting Eq.(\ref{eqn:pot2}) and denoting the angle $\theta-f$ as
$\varphi$, we have
\begin{eqnarray}
    \phi_{ml} &=& -{{GM_{\rm x}} \over a}
    \left\{{2 \over \pi^2} \int_0^\pi d(\Omega_B t) {a \over r_2}
    \cos(mf - l\Omega_B t) \right. \nonumber\\
    && \times \int_0^\pi d\varphi
    {{\cos m\varphi} \over (1+\beta^2-2\beta \cos \varphi)^{1/2}} \nonumber\\
    && \left. -{{\delta_{m1}} \over \pi} \int_0^\pi d(\Omega_B t) {a \over r_2}
    \beta \cos(mf - l\Omega_B t) \right\},
    \label{eqn:Fourier}
\end{eqnarray}
where $\beta = r/r_2$ and $\delta_{m1}$ is the Kronecker delta
function.

For each potential component $\phi_{ml}$, there can be outer and inner
Lindblad resonances at radii where $\Omega_p=\Omega \pm \kappa/m$ and
a corotation resonance at the radius where $\Omega_p=\Omega$. Here,
$\kappa$ is the epicyclic frequency, and here and hereafter the upper
and lower signs correspond to the outer Lindblad resonance (OLR) and
inner Lindblad resonance (ILR), respectively. The radii of these
resonances are given by
\begin{equation}
    r_{\rm LR} = \left( {{m \pm 1} \over l} \right)^{2/3}
                   (1+q)^{-1/3} a
    \label{eqn:rad_lr}
\end{equation}
and
\begin{equation}
    r_{\rm CR} = \left( {m \over l} \right)^{2/3}
                   (1+q)^{-1/3} a,
    \label{eqn:rad_cr}
\end{equation}
where $q=M_{\rm x}/M_{*}$.

For near-Keplerian discs, where $\Omega \sim \kappa \sim (GM_{*}/r^3)^{1/2}$,
Goldreich \& Tremaine's (1979; 1980) standard formula for torques
$T_{ml}$ at the outer Lindblad resonance (OLR) and inner Lindblad
resonance (ILR) is reduced to
\begin{equation}
    T_{ml} = \pm {{m(m \pm 1)\pi^2 \sigma (\lambda \mp 2m)^2
             \phi_{ml}^2}
             \over {3l^2 \Omega_B^2}},
    \label{eqn:tml_lr}
\end{equation}
where $\sigma$ is the surface density of the disc at the resonance
radius and $\lambda = (d \ln \phi_{ml}/d \ln r)_{\rm LR}$. Similarly,
the torque at the corotation resonance (CR) is written as
\begin{equation}
    T_{ml} = {{2m^3 \pi^2 \sigma \phi_{ml}^2}
             \over {3l^2 \Omega_B^2}}.
    \label{eqn:tml_cr}
\end{equation}
Note that angular momentum is removed from the disc at the ILRs,
whereas it is added to the disc at the OLRs and CRs.

In near Keplerian discs, the viscous torque formula derived by Lin \&
Papaloizou (1986) is written as
\begin{equation}
    T_{\rm vis} = 3 \pi \alpha GM_{*} \sigma r \left( {H \over r}
                  \right)^2
    \label{eqn:t_vis}
\end{equation}
[see also Artymowicz \& Lubow (1994)], where $\alpha$ is the
Shakura-Sunyaev viscosity parameter and $H$ is the vertical
scale-height of the disc. In our isothermal disc model (the parameters
used are listed in Table~\ref{tbl:params}), $H$ is given by
\begin{equation}
    {H \over r} = % 2.592 \cdot 10^{-2} \times 11.85^{1/2}
                  8.9 \cdot 10^{-2}
                  \left({r \over a} \right)^{1/2}.
    \label{eqn:h}
\end{equation}

The disc is truncated if the viscous torque is
smaller than the resonant torque. The criterion for the disc
truncation at a given resonance radius is, in general, written as
\begin{equation}
    T_{\rm vis} + \sum_{ml}(T_{ml})_{\rm ILR}
    +\sum_{ml}(T_{ml})_{\rm OLR}+\sum_{ml}(T_{ml})_{\rm CR} \le 0,
    \label{eqn:criterion}
\end{equation}
where the summation is taken over all combination of $(m,l)$ which
give the same resonance radius.
For example, at the $n$:1 resonance radius, at which $\Omega=n\Omega_{\rm B}$,
the summation is taken over all combinations of $(m, l)$ with
$l=n(m-1)$ for ILRs, $l=n(m+1)$ for OLRs, and $l=nm$ for CRs.
Actually,
criterion~(\ref{eqn:criterion}) is determined only by the viscous
torque and the torques from the ILRs of several lowest-order potential
components, because the torques from the ILRs dominate those from the
OLR and CR in circumstellar discs
and high-order potential components contribute little to the total
torque, even if the eccentricity of the orbit $e$ is not small
(Goldreich \& Tremaine 1980; Artymowicz \& Lubow 1994).

\begin{figure}
    \resizebox{\hsize}{!}{\includegraphics{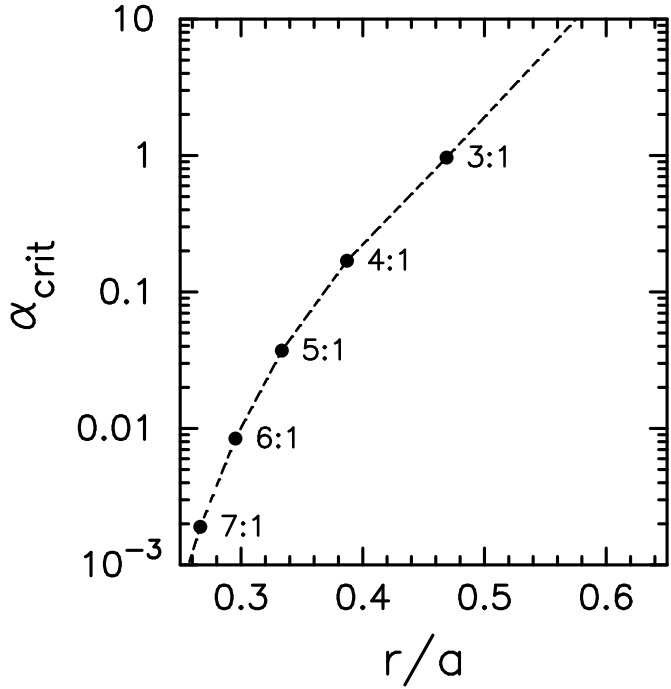}}
    \caption{Critical values of $\alpha$ at $n:1$ resonance
      radii. Annotated in the figure are the locations of the $n:1$
      commensurabilities of disc and binary orbital periods.}
    \label{fig:alpha_c}
\end{figure}

Criterion~(\ref{eqn:criterion}) at a given resonance is met for
$\alpha$ smaller than a critical value, $\alpha_{\rm crit}$. In
Fig.~\ref{fig:alpha_c}, we plot  $\alpha_{\rm crit}$ at $n:1$
resonance radii. The resonant torques at the $n:1$ radii are stronger
than those at radii with other period commensurabilities located
nearby. It is important to note that $\alpha_{\rm crit} \ga 1$ at the
2:1 and 3:1 resonance radii. The disc
is truncated at the 3:1 radius if $0.17 \la \alpha \la 0.97$ or at the
4:1 radius if $0.037 \la \alpha \la 0.17$. Note also that the Be-star
disc in 4U\,0115+63 cannot be in a steady state. The gas supplied by
the Be star will decrete outward and be accumulated in the outermost
part of the disc. This could contribute to make the disc dynamically
unstable, as will be discussed in Paper II.

It could be argued that the tidal truncation of the circumstellar disc
is an assumption implicit in our choice of a viscous decretion model
for the disc, but it holds irrespective of disc model as long as the
outflow velocity in the disc is very subsonic, as observed for many Be
stars. In order to show this, let us compare the timescale for disc
truncation, $\tau_{\rm trunc}$, with the drift timescale $\tau_{\rm
   drift}$. We can find $\tau_{\rm trunc}$ via the viscous timescale
$\tau_{\rm vis}$. Since the ratio of the viscous torque to the
resonant torque is $\alpha/\alpha_{\rm crit}$, we have
$\tau_{\rm trunc} \sim (\alpha/\alpha_{\rm crit})\tau_{\rm vis} \sim
\alpha_{\rm crit}^{-1}(\Delta r/H)^2\Omega^{-1}$. In deriving the
second similarity, we have used $\tau_{\rm vis} \sim (\Delta
r)^2/\alpha c_s H \sim (\Delta r)^2/\alpha H^2 \Omega$, where $c_s$ is
the sound speed and $\Delta r$ is a gap size (i.e., an interval
between the disc outer radius and the radius at which the neutron
star's gravity begins to dominate). On the other hand, the drift
timescale is written as
$\tau_{\rm drift} \sim \Delta r/v_r \sim {\cal M}_r^{-1}(\Delta r/H)
\Omega^{-1}$, where $v_r$ and ${\cal M}_r$ are the radial velocity and
Mach number, respectively. Consequently, we have
\begin{equation}
    {\tau_{\rm trunc} \over \tau_{\rm drift}}
    \sim {{\cal M}_r \over \alpha_{\rm crit}}
    {{\Delta r} \over H}.
    \label{eqn:timescale}
\end{equation}
As shown in Fig.~\ref{fig:alpha_c}, $\alpha_{\rm crit} \sim 1$ at the
3:1 radius. From Eq.(\ref{eqn:h}) with $r/a \sim 0.5$, we expect
$\Delta r/H = \Delta r/r/(H/r) \sim O(1)$ for the disc of V635~Cas.
Therefore, we conclude that the disc truncation timescale is much shorter
than the drift timescale if the outflow is very subsonic.
Note that this conclusion holds for Be/X-ray binaries in general,
because all Be/X-ray binaries have similar disc parameters
and the orbital parameters of these systems do not range very widely.

\subsection{Disc structure}
\label{sec:uptbdisc}

As mentioned above, the disc around V635~Cas cannot be
steady. Nevertheless, it is instructive to study the steady disc
structure as a model for the inner disc and/or the disc in the initial
formation epoch. In this section, we assume the disc to be axisymmetric
and azimuthally average the potential given by Eq.~(\ref{eqn:pot1}).

Okazaki (1997; see also Okazaki 2000b) showed that the equations
which determine the velocity field in an isothermal decretion disc can
be reduced to
\begin{equation}
     \left( V_r - {c_{\rm s}^2 \over V_r} \right)
        {{{\rm d}V_r} \over {{\rm d}r}}
        = -{{GM_{*}}\over{r^2}} + F_{\rm rad}
        + {\ell^2 \over r^3} + {5 \over 2}{c_{\rm s}^2 \over r}
        \label{eqn:windeq1}
\end{equation}
and
\begin{equation}
     \ell = \ell (R_*) + \alpha c_{\rm s}^2
               \left[{R_* \over V_r(R_*)}-{r \over V_r}\right],
        \label{eqn:windeq2}
\end{equation}
where $\ell=rV_\phi$ is the specific angular momentum,
$F_{\rm rad}$ is the vertically averaged radiative force,
and $V_r$ and $V_\phi$ are the radial and azimuthal components
of the vertically averaged velocity, respectively.
Eqs.~(\ref{eqn:windeq1}) and (\ref{eqn:windeq2}) show that the viscous
decretion disc is an equatorial wind, in which material is slowly
accelerated outward by the pressure force.
 
Since the radial flow in Be-star discs is considered to be subsonic,
the radiative  force would arise not from the optically-thick strong
lines but from  an ensemble of optically-thin weak lines.
Hence, for the radiative force, we adopt the parametric form
proposed by Chen \& Marlborough (1994).
\begin{equation}
    F_{\rm rad} \simeq \frac{GM_{*}}{r^2}
              \eta \left( {r \over R_*} \right)^\epsilon,
    \label{eqn:frad}
\end{equation}
where $\eta$ and $\epsilon$ are parameters which characterize
the force due to the ensemble of optically-thin lines.
 
Given the radial velocity component at the stellar surface, the flow
structure in the disc is obtained by solving Eqs.~(\ref{eqn:windeq1})
and (\ref{eqn:windeq2}) numerically. The surface density distribution
is then obtained from the equation of mass conservation.
In Fig.~\ref{fig:decretion}, we show the structure of
the isothermal decretion disc around V635~Cas with
$\alpha=0.1$, $V_r(R_*)/c_{\rm s} = 10^{-3}$, and the outer
radius at the 4:1 resonance. The parameters characterizing
the radiative force, $\eta$ and $\epsilon$, are chosen
so that the fundamental $m=1$ mode in the disc has the period
corresponding to the observed line profile variabilities
(see Paper~II).

The solution shown in Fig.~\ref{fig:decretion} is not unique. We found
that, for $\alpha=0.1$, there is a smooth solution for
$V_r(R_*)/c_{\rm s} < 3 \times 10^{-3}$. Any flow with $\alpha=0.1$ and
$V_r(R_*)/c_{\rm s} \ga 3 \times 10^{-3}$, however, has a spiral
critical point and does not go smoothly from the stellar surface to
the disc outer radius. Thus, the radial flow in a decretion disc must
be highly subsonic near the star. Such a flow satisfies the disc
truncation condition, $\tau_{\rm trunc}/\tau_{\rm drift}<1$, discussed
in $\S$\,\ref{sec:trunc}.
 
\begin{figure}
    \resizebox{\hsize}{!}{\includegraphics{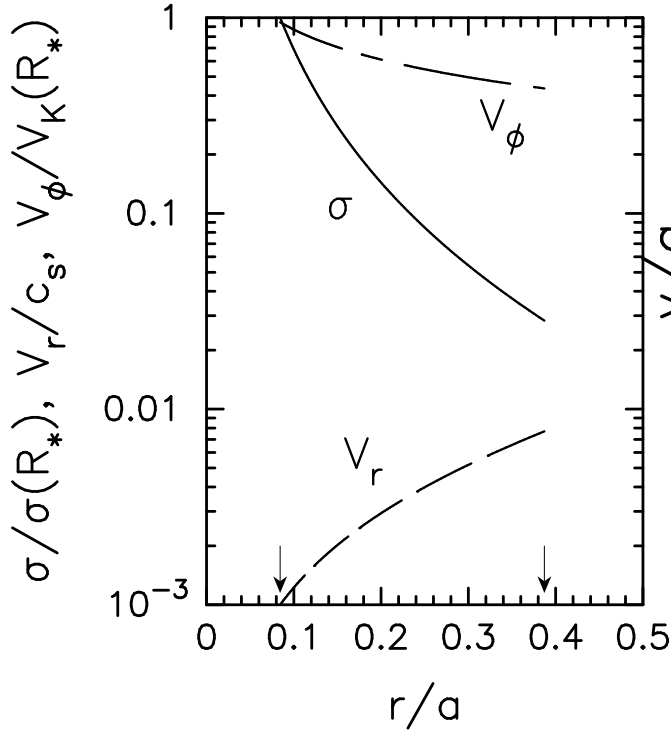}}
    \caption{Structure of the viscous decretion disc around
      V635~Cas with $\alpha=0.1$ and $V_r(R_*)/c_{\rm s} =
      10^{-3}$.
      The radiative force in the form of
      $\eta(r/R_*)^\epsilon \times GM_{*}/r^2$ with
      $(\eta, \epsilon)=(0.1, 0.1)$ is included.
      Solid, dashed, and dash-dotted lines denote
      $\sigma/\sigma(R_*)$, $V_r/c_{\rm s}$, and
      $V_\phi/V_{\rm K}(R_*)$, respectively,
      where $V_{\rm K}=(GM_{*}/r)^{1/2}$ is the Keplerian velocity.
      Two arrows denote the inner radius at $r=R_{*}$
      and the outer disc radius located at the 4:1 resonance radius.}
    \label{fig:decretion}
\end{figure}

\section{Discussion}
\label{sec:discuss}

The truncation of the disc by the tidal/resonant interaction with the
neutron star could explain many of the observed properties of V635
Cas. In N97, it was shown that large changes in the infrared
magnitudes were not associated with significant changes in the
associated colours. According to the models of Dougherty et
al. (1994), this should correspond to a very small disc with a very
high density - such that it is optically thick at all wavelengths for
continuum radiation. The formation of such a disc is easy to explain
if tidal truncation forces outflowing material to fall back on to the
inner regions of the disc. It will also explain why the observed
correlation between the strength of H$\alpha$ and the circumstellar reddening
observed in Be stars (Fabregat \& Reglero 1990) does not seem to hold for
Be/X-ray binaries - the disc could become very dense and the circumstellar
reddening would not be measuring the size of the disc. Finally it would
explain the correlation between the maximum strength of H$\alpha$ and
orbital period in Be/X-ray binaries found by Reig et al. (1997).

\begin{figure}
\begin{picture}(250,250)
\put(0,0){\includegraphics{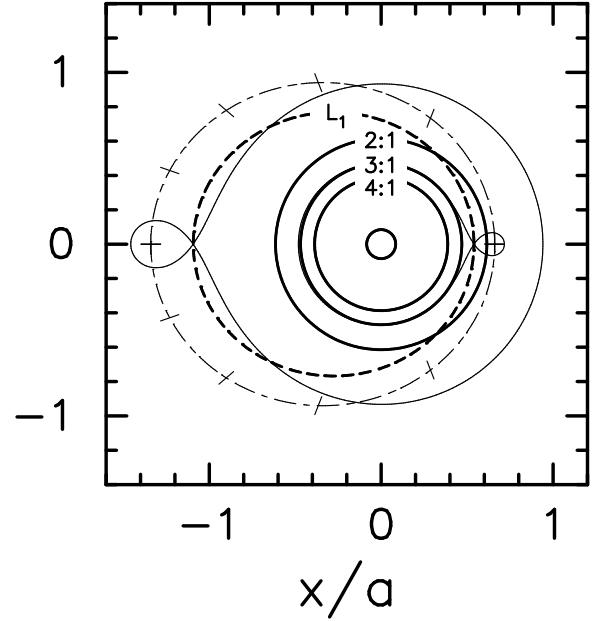}}
\end{picture}
\caption{A description of the orbital model adopted for 4U\,0115+63,
   in the reference system centred on the Be star. The dash-dotted
   line represents the orbit of the neutron star. The thick dashed
   line represents the position of the first Lagrangian point ($L_1$)
   around the orbit. The solid thin lines are the effective Roche
   lobes of the two stars at apastron and periastron (the position
   of the neutron star is marked with a cross). The labelled thick
   solid lines are the locations of the $n:1$ commensurabilities of
   disc and binary orbital periods. Note that the radius of the 3:1
   commensurability is smaller than the radius of the $L_1$ point at
   periastron.}
\label{fig:orbit} 
\end{figure}

The reality of this correlation has been questioned by Apparao (1998),
who claims that the EW of H$\alpha$ does not represent an appropriate
measurement of the size of the disc and that the actual luminosity 
radiated in H$\alpha$ does not show any correlation with the size of
the orbit. We agree that the EW of H$\alpha$ at a given
time does certainly not represent a measurement of the size of the envelope.
Many other factors such as the optical properties of the envelope, 
which seem to be highly variable, or an asymmetric density distribution
in the disc (Negueruela et al. 1998; see also Paper II) affect the
observed EW. However, the $maximum$ EW of H$\alpha$ ever observed may 
provide a comparative estimate of the maximum size of an envelope if the
source has been consistently monitored for a sufficiently long time -- for 
example in the case of V635~Cas, where the typical timescales of 
variability  are of the order of a few months (Paper II), ten years
of monitoring provide a good estimate of the EW of H$\alpha$ at
maximum.

Given that we expect the viscosity in the disc to be of the order of
$\alpha \la 0.1$ as in accretion discs (e.g., Blondin 2000; Matsumoto
1999), the disc around V635~Cas is likely to be  
truncated at the 4:1 resonance. Even if the viscosity is higher and becomes 
$\approx 1$, truncation still occurs at the 3:1 resonance. In 
Fig.~\ref{fig:orbit}, it can be seen that even the 3:1 resonance is 
significantly smaller than the radius of the first Lagrangian point
(even when the neutron star is at periastron).
Here, the potential $\psi$ describing the 
effects of the gravitational and centrifugal forces on the motion of
test particles orbiting the Be star is given by
\begin{equation}
   \psi (r, \theta, z) = \Phi (r, \theta, z)
   -{1 \over 2} \Omega^2(r) r^2,
   \label{eqn:roche}
\end{equation}
where $\Phi$ is the potential defined by Eq.~(\ref{eqn:pot1}).
As a consequence, the neutron star will not be able to
accrete a significant amount of material from the circumstellar disc.
This fact explains the lack of series of Type I outbursts from 4U\,0115+63.
In order to have Type~I outbursts in this
system, we will need a large amplitude $m=1$ mode to make the disc
eccentric and cause Roche lobe overflow toward the neutron star.

Accepting that the disc of Be/X-ray binaries is tidally truncated 
naturally explains why Be/X-ray binaries spend most of their time in 
quiescence. 
In general, the neutron star cannot accrete from the dense regions of 
the disc. The low-luminosity X-ray activity observed in systems such
as A\,0535+262 (Motch et al. 1991) would then be due to the accretion of
some residual low-density outflow. In the case of V635~Cas, this material
is not accreted because of the propeller mechanism. If we accept this
picture, it is precisely the very existence of X-ray outbursts that needs
to be explained. This will be addressed in Paper II.

\section{Conclusions}

By observing the source during a disc-loss episode, we have been 
able to determine the stellar parameters of 
V635~Cas. The object is classified as a B0.2Ve star at a distance of 
$7-8$ kpc. Both the mass function and the estimated $v\sin i$
indicate a moderate inclination for the orbital and equatorial planes.
The derived distance implies that the source can radiate
close to the Eddington luminosity for a neutron star during bright outbursts. 

With the newly determined parameters, we have constructed a model for
4U\,0115+63. Based on the viscous decretion disc model for Be stars, we
have numerically solved a criterion for tidal truncation and found
that the disc surrounding V635~Cas must be truncated at a resonance
radius depending on the viscosity parameter and cannot be 
in a steady state. Although we have adopted a particular disc model, our
conclusion is robust as long as the outflow in the disc is subsonic, 
a hypothesis supported by several observational facts. Under normal
conditions, the neutron star cannot accrete enough material to overcome
the centrifugal barrier and switch on the X-ray emission.

\section*{Acknowledgements}

 This work would not be possible without
the ING service programme. Special thanks to the service manager 
Steve Smartt and to Don Pollacco for their interest
in the monitoring of V635~Cas. The WHT and JKT are operated on the 
island of La Palma by the Royal Greenwich Observatory in the Spanish 
Observatorio del Roque de Los Muchachos of the Instituto de
Astrof\'{\i}sica de Canarias. This research has made use of the Simbad 
data base, operated at CDS, Strasbourg, France. IN would like to 
thank Malcolm Coe for drawing his attention to this source. IN has
benefited from many helpful discussions with Drs. John Porter and
Iain Steele. B. Skouza proved to be inspirational.

We thank the referee, Dr. W. Hummel, for his helpful criticism, which
contributed to improve the paper notably.
Data reduction was mainly carried out using the Liverpool
John Moores University {\em Starlink} node, which is funded by PPARC. 
During part of this work, IN was supported by a PPARC postdoctoral 
fellowship. At present he holds an ESA external fellowship.

\end{document}